\documentclass[twocolumn,prl,aps,floatfix,groupedaddress]{revtex4}
\usepackage{amsmath}
\usepackage{graphicx}
\usepackage{amssymb}

\newcommand{\e}[1]{\mathrm{e}^{#1}}
\renewcommand{\th}[1]{\theta_3\left(0,\mathrm{e}^{#1}\right)}

\begin{document}

\title{Interacting bosons in one dimension and Luttinger liquid theory} 

\author{Adrian Del Maestro}
\affiliation{Department of Physics and Astronomy, University of British Columbia,
Vancouver, British Columbia V6T 1Z1, Canada}

\author{Ian Affleck}
\affiliation{Department of Physics and Astronomy, University of British Columbia,
Vancouver, British Columbia V6T 1Z1, Canada}

\begin{abstract}
Harmonically trapped ultra-cold atoms and ${}^4$He in nanopores provide new experimental
realizations of bosons in one dimension, motivating the search for a more complete theoretical
understanding of their low energy properties.  Worm algorithm path integral quantum Monte Carlo
results for interacting bosons restricted to the one dimensional continuum are compared to the
finite temperature and system size predictions of Luttinger liquid theory. For large system sizes at
low temperature, excellent agreement is obtained after including the leading irrelevant interactions
in the Hamiltonian which are determined explicitly.
\end{abstract}

\maketitle
Luttinger liquid (LL) theory \cite{haldane} provides a universal description of interacting fermions
or bosons at sufficiently low energies in one dimension (1D).  Recently, exciting new possibilities
for experimental realizations of Luttinger liquids have appeared, involving ultra-cold atoms
in cigar-shaped traps \cite{greiner},  screw dislocations in solid ${}^4$He \cite{screw} and
helium-4 confined to flow in nanopores \cite{gervais}.  In these systems, a translationally
invariant model of interacting bosons in free space may be a good starting point.  While there
have been numerous numerical studies of 1D fermion models on lattices using exact diagonalization,
Monte Carlo and Density Matrix Renormalization Group methods, numerical results on free space
interacting bosons at non-zero temperature, $T$, are much rarer.  Exact studies in the continuum may
provide new insights, specifically on issues of dimensional crossover in nanopores.  Zero
temperature variational Monte Carlo calculations for the 1D case were reported in Ref.~\cite{dqmc}
and finite $T$ worm algorithm path integral Monte Carlo (WA-PIMC) simulations for a screw
dislocation by Boninsegni \emph{et al.} \cite{screw} have claimed the observation of LL behavior.
In order to systematically explore the regime of energies and pore lengths where LL behavior may
occur we have performed WA-PIMC simulations on the $N$-particle Hamiltonian:
\begin{equation}
H = -\frac{1}{2m} \sum_{i=1}^{N} \nabla_i^2 + \frac{1}{2}\sum_{i,j=1}^{N}V(|\vec{r}_i - \vec{r}_j|)
\label{eq:hamMicroscopic}
\end{equation}
in 1D with periodic boundary conditions on an interval of length $L$ in angstroms and we will work
in units where $\hbar=k_\mathrm{B}=1$.  The WA-PIMC method, recently introduced by Boninsegni
\emph{et al.} \cite{worm} extends the original PIMC algorithm of Ceperley \cite{ceperleyReview} to
include configurations of the single particle Matsubara Green function, allowing for intermediate
particle trajectories which are not periodic in imaginary time. The inclusion of such trajectories
yields an efficient and robust grand canonical quantum Monte Carlo (QMC) technique that accurately
incorporate complete quantum statistics and provides exact and unbiased estimations of many physical
observables at finite temperature.  In the WA-PIMC simulations performed here, the short range
repulsive interaction $V(r) = (g/\sqrt{\pi}a) \mathrm{e}^{-r^2/4a^2}$ is chosen for convenience to
be Gaussian,  with integrated strength $2g$ and spatial extent $a$. The numerical values of all
microscopic parameters were optimized to ensure an experimentally relevant and efficient grand
canonical simulation at low energies where the temperature is much smaller than both the kinetic
($E_K/N$) and potential ($E_V/N$) energy per particle.  To obtain $E_K/N \sim E_V/N \sim 5$~K we
have fixed the chemical potential at $\mu=24$~K with $2g=20$~K and the interaction width
$a\simeq0.03$~\AA~ to be much less than the resulting inter-particle separation, $1/\rho_0\simeq
0.67$~\AA~ for particles of mass $m=0.25$~\AA${}^{-2}$K${}^{-1}$.

Luttinger liquid theory uses a low energy effective harmonic Hamiltonian to capture the quantum
hydrodynamics of a microscopic 1D system. This is accomplished in terms of two bosonic fields,
$\theta(x)$ and $\phi(x)$ representing the density and phase oscillations of a particle field
operator
\begin{equation}
H_\mathrm{LL} = \frac{1}{2\pi}\int_0^L dx \left[v_J \left(\partial_x \phi\right)^2 + v_N
\left(\partial_x \theta\right)^2\right],
\label{eq:LLHam0}
\end{equation}
where the two velocities $v_J$ and $v_N$ are directly related to the microscopic details of the
underlying high energy model.  If the system of interest exhibits Galilean invariance, $v_J = \pi
\rho_0/m$ and $v_N = 1/(\pi \rho_0^2 \kappa)$ where $\rho_0=N_0/L$ is the mean number density and
$\kappa$ is the adiabatic compressibility in the limit $L\to \infty$, $T\to 0$ \cite{haldane}. 

In this study we find that that the mean number of particles at finite temperature $\langle N
\rangle$ exhibits corrections to scaling that are \emph{not} captured by Eq.~(\ref{eq:LLHam0}).
Instead, through a detailed analysis of the super and normal fluid components of the one dimensional
repulsive Bose gas, we argue that the observed deviations from scaling result from higher order
``irrelevant'' operators that should be included in the low energy effective Hamiltonian.

Although the QMC performed here allows access to a large number of properties of the microscopic
system, in order to study the applicability of the effective model in Eq.~(\ref{eq:LLHam0}) it will
be enough to focus on the probability distributions for number and phase fluctuations. Within the LL
theory, these are most easily derived by performing a mode expansion of $\theta(x)$ and $\phi(x)$
for periodic boundary conditions indexed by wavevector $q=2\pi n/L$ (see Ref.~\cite{haldane}). The
grand partition function can then be written as 
\begin{align}
\mathcal{Z} &= \e{-\epsilon_0 L/T} \e{\pi v / 6LT} \sum_{N}\e{-\pi v_N N^2/2LT + \mu N/T} \nonumber \\
& \qquad \times \sum_{J}\e{-\pi v_J J^2/2LT} \prod_{n\ne0}\left(1-\e{-2 \pi v |n|/LT}\right)^{-1}
\label{eq:Z0}
\end{align}
where $J$ is an even integer indexing topological excitations (winding) of the phase field $\phi$,
$\epsilon_0$ is the non-universal ground state energy per unit length and $v$ is the phonon velocity
given by the algebraic mean of $v_J$ and $v_N$: $v = \sqrt{v_J v_N}$. By tracing out winding and
phonon modes, which cannot affect the density, we immediately arrive at an expression for the
particle number probability distribution
\begin{equation}
P(N) = \frac{\e{-\frac{\pi v_N}{2LT}(N-N_0)^2}}{\th{-\pi v_N/2LT}},
\label{eq:P0N}
\end{equation}
where  $\theta_3(z,q)$ is a Jacoby Theta function of the third kind. 
%
\begin{figure}[t]
\centering
\includegraphics[clip,width=1\columnwidth]{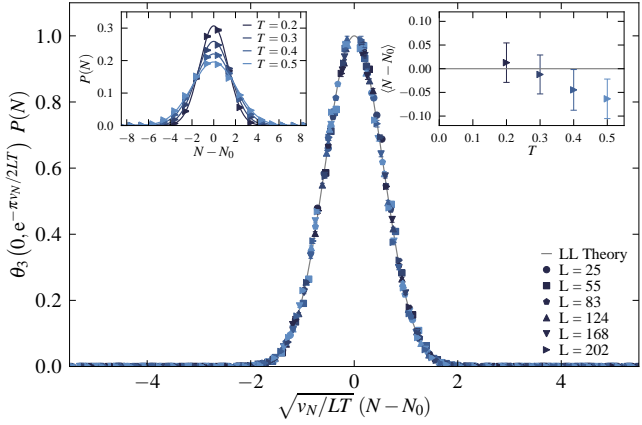}
\caption{\label{fig:particle} (Color online) QMC data (symbols) combined with Luttinger liquid
predictions (sold lines) for the particle number probability distribution at fixed system size
(upper left inset), scaling of the particle number probability distribution (main panel) and the
temperature dependence of the mean number of particles (upper right inset) measured with respect to
the ground state value $N_0 = \rho_0 L$.}
\end{figure}
%
An immediate consequence of this result is that LL theory predicts that the average number of
particles exhibits no temperature dependence, $\langle N-N_0\rangle = 0$ and it is on this red
herring that we will focus our attention below.  An equivalent expression for $P(J)$ can be derived
in the same manner. However, it will be more useful to work with a dual coordinate for $J$ known as
the winding number $W$ which is easily measured in the QMC \cite{ceperleyWinding} and is related to
the wrapping of imaginary time particle trajectories around the physical boundaries of the sample
\begin{equation}
P(W)= \frac{\e{-\frac{\pi L T}{2 v_J}W^2}}{\th{-\pi L T/ 2 v_J}}.
\label{eq:P0W}
\end{equation}
%
\begin{figure}[t]
\centering
\includegraphics[clip,width=1\columnwidth]{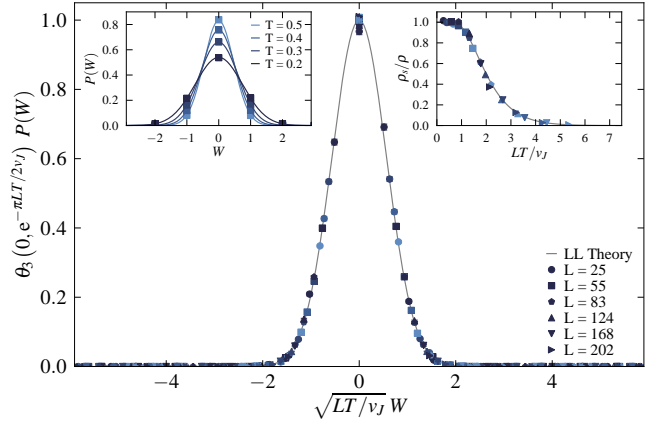}
\caption{\label{fig:winding} (Color online) QMC data (symbols) combined with Luttinger liquid
predictions (sold lines) for the winding number probability distribution at fixed system size (upper
left inset), scaling of the winding number probability distribution (main panel) and the superfluid
fraction as a function of the dimensionless scaling variable $LT/v_J$ (upper right inset).}
\end{figure}
%
The strange \emph{inverse} Boltzmann form of this distribution can be understood by noting that in
one dimension, the superfluid density is proportional to the second moment of the winding number
distribution \cite{ceperleyWinding} and it is only when fluctuations of the phase field $\phi$ are
suppressed (phase coherence with $\langle J \rangle \sim 0$) that the system will acquire a finite
superfluid response. As a consequence of Eq.~(\ref{eq:P0W}), the superfluid fraction will be a pure
scaling function of $v_J/LT$ given by \cite{dalong}
\begin{equation}
\frac{\rho_s}{\rho} = 1 - \frac{\pi v_J}{LT} 
\left|\frac{\theta_3''(0,\e{-2\pi v_J/LT})}{\th{-2\pi v_J/LT}}\right|
\label{eq:rhosScale}
\end{equation}
where $\theta_j''(z,q) \equiv \partial^2_z \theta_j(z,q)$.

The above theoretical predictions can be verified by investigating the particle and winding number
probability distributions measured in the QMC for a range of temperatures and system sizes.  It is
crucial to recognize that the \emph{parameters} of LL theory, $v_J$ and $v_N$, have no temperature
or finite size dependence and depend only on the microscopic details of the high energy theory in
Eq.~(\ref{eq:hamMicroscopic}).  Both $P(N)$ and $P(W)$ are scaling functions of $LT/v_{J,N}$, and
fits of numerical data for an individual system size at fixed temperature \emph{must} produce values
of $v_J$ and $v_N$ that work equally well at all $L$ and $T$ provided the system is in a regime
where the LL theory of Eq.~(\ref{eq:LLHam0}) is applicable.  Figs.~\ref{fig:particle} and
\ref{fig:winding} present a summary of our QMC data for $L=25-202$~\AA~ and $T=0.2-0.5$~K.  The
insets in the upper left hand corners of these figures show the result of fitting to
Eqs.~(\ref{eq:P0N}) and (\ref{eq:P0W}) yielding $v_J = 18.88(2)$~\AA K and $v_N = 7.7(3)$~\AA K
which combine to give a Luttinger parameter of $K = \sqrt{v_N/v_J} = 0.64(4)$ where the number in
brackets gives the uncertainty in the final digit. As mentioned earlier, the presence of Galilean
invariance relates $v_J$ and $\rho_0$ and we find the zero temperature equilibrium density to be
$\rho_0 = 1.5028(2)$~\AA${}^{-1}$.  The main panel in both plots shows data collapse over all system
sizes and temperatures measured that appears to be consistent with the scaling predictions of LL
theory.  However, a closer look at the average number of particles as a function of temperature for
$L=202$~\AA~ (Fig.~\ref{fig:particle} upper right inset) clearly shows that as the temperature is
increased at fixed chemical potential, the mean particle number is decreasing, in stark
contradiction with the prediction of Eq.~(\ref{eq:P0N}).  Conversely, scaling in the winding number
sector shows no such deviations and the computed superfluid fraction $\rho_s/\rho$ plotted as a
function of the dimensionless scaling variable $LT/v_J$ (Fig.~\ref{fig:winding} upper right inset)
is indistinguishable from the LL prediction of Eq.~(\ref{eq:rhosScale}).

At sufficiently low $T$ and large $L$, corrections to scaling should be described by the leading
irrelevant interactions added to the Hamiltonian of Eq.~(\ref{eq:LLHam0}). Recently, such
corrections were shown to lead to qualitative modifications of the spectral function for fermions
\cite{imambekov}. Assuming that all interactions are short range, correcting terms come from an
expansion of the kinetic energy in Eq.~(\ref{eq:hamMicroscopic}) and are related to band curvature
effects.  The lowest order correction to the LL Hamiltonian containing three derivatives, consists
of two operators
\begin{equation}
H^\prime = \frac{1}{2\pi^2 \rho_0}\int_0^L dx \left[ v_J\left(\partial_x \phi\right)^2\partial_x
\theta + \lambda v_N\left(\partial_x\theta\right)^3\right]
\label{eq:LLHamCorr}
\end{equation}
where the coefficient of the first term is constrained by Galilean invariance.  The form of
$H^\prime$ could also have been inferred on phenomenological grounds alone, as these two terms are
the only dimension three operators that are allowed by parity under which $\theta \to -\theta$,
$\phi \to \phi$ and $x \to -x$.  The dimensionless factor $\lambda$ in the second coefficient can be
determined \cite{pereira} by noting that in the ground state, if we shift the chemical potential by
an infinitesimal amount $\mu \to \mu + \delta \mu$ there will be a corresponding shift in the
density $\rho_0 \to \rho_0 + \delta \rho_0$ governed by the thermodynamic relation $\delta \rho_0 =
\rho_0^2 \kappa \delta \mu$. Keeping terms to $O(\delta\mu)$ in an expansion of the ground state
Gibbs free energy for $H_\mathrm{LL} + H^\prime$ we find $\lambda = (\pi \rho_0/3) \partial_\mu v_N
= (\rho_0/3)\partial_\mu (\rho_0^2 \kappa)^{-1}$.

The influence of the corrections on thermodynamic quantities can be most easily understood by again
performing a mode expansion of the bosonic field $\theta$ and $\phi$.  Gradients of $\theta$ are
related to fluctuations of the particle number away from its mean value $\partial_x \theta \sim
N-N_0$ and thus the addition of linear and cubic terms will cause both a shift and skew in the
particle number probability distribution $P(N)$ in Eq.~(\ref{eq:P0N}) without changing its width.
On the other hand, the only corrections at this order to the winding number distribution $P(W)$
would have to come from the first term in Eq.~(\ref{eq:LLHamCorr}), but due to the linear power of
$\partial_x \theta$ which accompanies it as a multiplicative factor, any trace over the number of
particles when computing the grand partition function would cause its effects to average to zero.
This is exactly the qualitative behavior we observed from the analysis of our numerical results.  

In order to quantify these arguments, we may calculate the deviation in the mean number of particles
$\langle N - N_0 \rangle$ in a perturbative expansion of $H_\mathrm{LL} + H^\prime$ in the inverse
system size $1/L$ \cite{sirker}.  The resulting corrected deviation in the mean number of particles
is given by 
\begin{equation}
\langle N - N_0 \rangle = -\frac{1}{N_0}
\Phi_N\left(\frac{LT}{v_N},\frac{LT}{v_J},\lambda\right)
\label{eq:NShift}
\end{equation}
where
\begin{multline}
\Phi_N(x,y,\lambda) = \frac{2x}{\pi} \left \{ 
\lambda\left[ \left( x\partial_x \ln \th{-\pi/2x} \right)^2 \right. \right. \\
\left.\left. +\; x^2\partial^2_x \ln \th{-\pi/2x} + 2 x\partial_x \ln \th{-\pi/2x} \right] \right. \\
\left. - \left[\frac{1}{2}(3\lambda + 1) \left(y\partial_x + x\partial_y \right)\ln \eta
\left(i\sqrt{xy}\right) \right. \right. \\
\left. \left. -\; y\partial_y \ln \th{-2\pi/y} \right] x\partial_x \ln \th{-\pi/2x} \right\}
\label{eq:PhiN}
\end{multline}
is a universal scaling function with $\eta(ix)$ being the Dedekind eta function. This rather
complicated looking expression has a simple asymptotic form in the thermodynamic limit where
$LT/v_{J,N} \to \infty$
\begin{equation}
\Phi_N\left(\frac{LT}{v_N}\to\infty,\frac{LT}{v_J}\to\infty,\lambda \right) \to
\frac{K}{12}\left( 3\lambda + 1\right)\left(\frac{LT}{v_N}\right)^2
\label{eq:PhiNTL}
\end{equation}
and $K=\sqrt{v_N/v_J}$ is the Luttinger liquid parameter.  It is now immediately clear that when
Eq.~(\ref{eq:PhiNTL}) is combined with Eq.~(\ref{eq:NShift}) a temperature dependent correction to
the mean density of particles will persist, even in the thermodynamic limit  
\begin{equation}
\langle\rho-\rho_0\rangle \to -\frac{K}{12\rho_0 v_N^2}\left( 3\lambda + 1 \right)\, T^2.
\label{eq:rhoT}
\end{equation} 
Indeed, from simple thermodynamic arguments one expects $\rho = -(1/L)\partial_\mu G$ where $G$, the
Gibbs free energy, can be easily calculated for a harmonic Luttinger liquid from Eq.~(\ref{eq:Z0})
to be $G = \epsilon_0 L - \mu \rho_0 L - (\pi L/6 v)T^2$  in the limit $LT/v \to \infty$.  When
performing the partial derivative of $G$ with respect to the chemical potential we recover the
asymptotic value in Eq.~(\ref{eq:rhoT}).

\begin{figure}[t]
\centering
\includegraphics[clip,width=1\columnwidth]{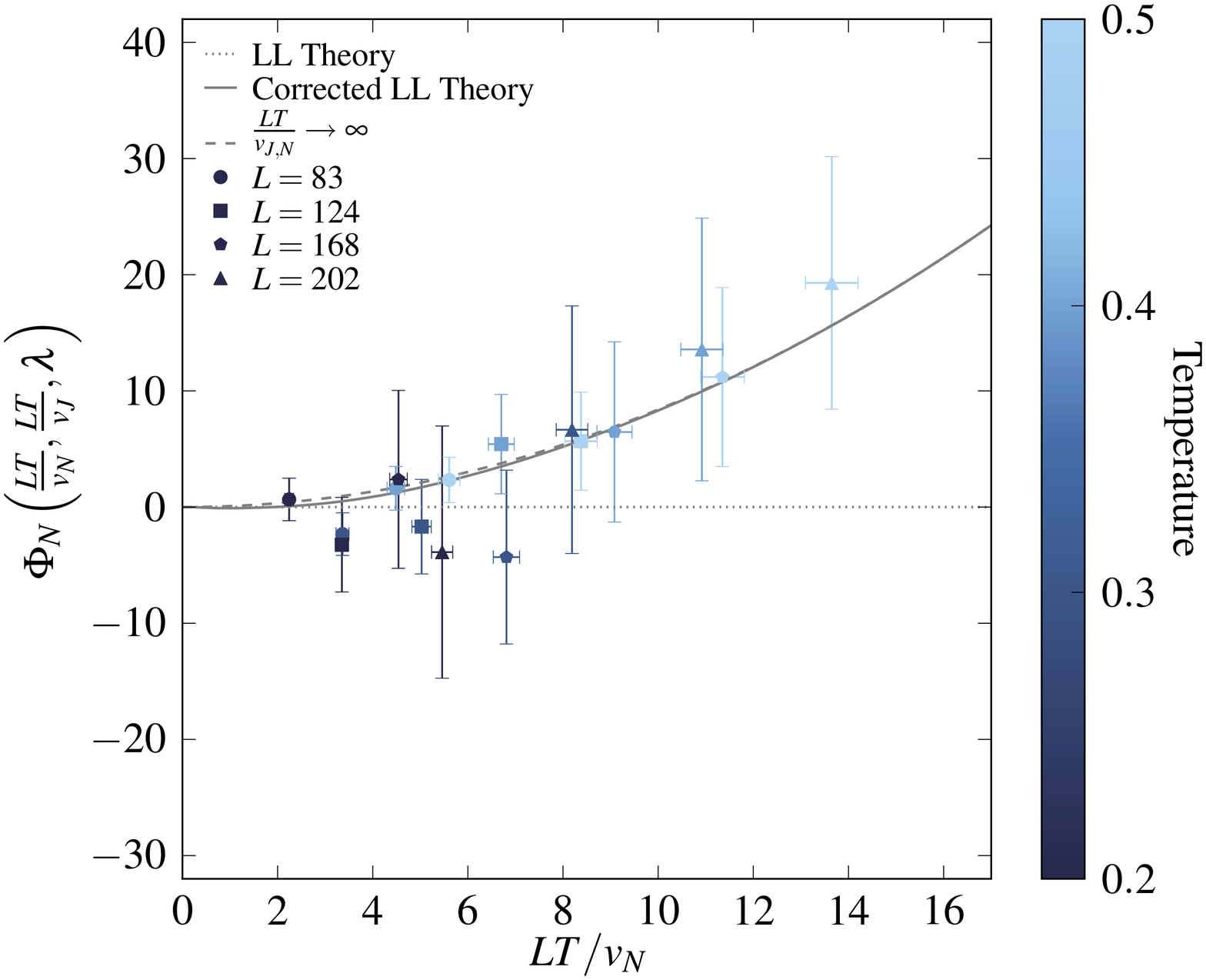}
\caption{\label{fig:numberScaling} (Color online) QMC data (symbols) combined with predictions
from the harmonic (dotted line) and corrected anharmonic (solid line) Luttinger liquid theory for
the universal scaling function $\Phi_N$. The dashed line shows the asymptotic result in the
thermodynamic limit from Eq.~(\ref{eq:PhiNTL}).}
\end{figure}

We are know in a position to test how well the extended Luttinger liquid Hamiltonian $H_\mathrm{LL} +
H^{\prime}$ captures these effects by comparing the scaling function $\Phi_N$ with our numerical
data.  The result is shown in Fig.~\ref{fig:numberScaling} where we have extracted the value of
%
%
$\lambda = 0.19(4)$ by re-fitting a perturbatively corrected number probability distribution to the
QMC results and used the previously extracted values of $v_{J,N}$.  The agreement between the
numerical QMC data (symbols) and the prediction of Eq.~(\ref{eq:PhiN}) (solid line) is found to be
statistically significant with a reduced maximum likelihood estimator of $\chi^2 \simeq 0.9$.  Large
uncertainties are unavoidable as there are inherent stochastic errors in the average density of
particles measured in the Monte Carlo that are magnified on this scale. Poorer agreement at low
temperatures is a reflection of the computational difficulty of performing ergodic numerical
simulations for nearly integrable systems \cite{kinoshita}.  

It is well known that an analytical solution of the delta-function interacting Bose gas can be
obtained via Bethe Ansatz (BA) \cite{liebLiniger} where the density dependence of the phonon
velocity $v$ can be extracted from an analysis of the linear coefficient of the long wavelength
dispersion relation \cite{lieb}.  It seems prudent to place our numerical data in this context and
for the mass, interaction strength and chemical potential used in the simulations, the
$T\to0,L\to\infty$ solution gives a Luttinger parameter of $K_{BA} \simeq 0.6299$ and cubic operator
coefficient $\lambda_{BA} \simeq 0.1272$.  The numbers extracted here of $K=0.64(4)$ and $\lambda =
0.19(4)$ agree relatively well within error-bars, but their systematically larger values are related
to the finite interaction width $a$ employed in the QMC.  This trend can be quantified by performing
the simulation for increasingly long-range interactions (although still requiring that $a\rho_0 \ll
1$) and for $a\rho_0 \simeq 0.06$ we find $K = 0.75(2)$ pushing us towards a regime with enhanced
charge density wave type order.

In addition to allowing for the study of finite range interactions, Monte Carlo methods can also
provide details on correlation functions that are not accessible via Bethe Ansatz \cite{dqmc}.
Unbiased measurements of the pair correlation function and single body density matrix computed in
the Monte Carlo are fully consistent with the predictions of LL theory and will be reported on
elsewhere \cite{dalong}.

In conclusion, we have performed large scale grand canonical worm algorithm path integral Monte
Carlo simulations of one dimensional repulsive soft-core bosons in the continuum at fixed chemical
potential.  We have shown that the finite size and temperature scaling behavior of the superfluid
density can be fully understood in terms of the low energy effective harmonic Luttinger liquid
theory provided the temperature is sufficiently small when compared to both the kinetic and
potential energy per particle and the system is large enough to overcome the effects of any finite
size gaps.  However, we have argued the temperature dependence of the mean particle density, a
quantity that can be easily measured in experiments on low dimensional bosonic systems, exhibits
corrections to scaling that can only be adequately accounted for by extending the theory to include
leading order irrelevant operators.

It is a pleasure to acknowledge R.~Melko, G.~Gervais, M.~Boninsegni and N.~Prokov'ev for many
fruitful and interesting discussions on various aspects of this study.  This work was made possible
through support from CIfAR (IA), NSERC (AD and IA) and the National Resource Allocation Committee of
Compute Canada with computational resources being provided by WESTGRID and SHARCNET.

\end{document}